\documentclass[preprint,journal]{IEEEtran}
\IEEEoverridecommandlockouts
\usepackage{tikz}
\usetikzlibrary{intersections,shapes,arrows,shadows,positioning ,arrows.meta,shadows.blur, matrix, calc, patterns, fit,automata}
\tikzset{rectangle, 
rounded corners =5pt,
minimum width =50pt, 
minimum height =20pt, 
inner sep=5pt, 
draw=blue }
\usepackage{tcolorbox}
\usepackage{bm}
\usepackage{algorithmic}
\usepackage{amsfonts, amsthm}
\newtheorem{assumption}{Assumption}
\newtheorem{theorem}{Theorem}
\newtheorem{corollary}{Corollary}

\usepackage{hyperref}
\newcommand{\mydoi}{10.1109/ICAIBD57115.2023.10206080} 
\usepackage{graphicx}

\usepackage{fancyhdr}

\ifCLASSINFOpdf
\else
\fi
\usepackage[cmex10]{amsmath}
\usepackage{cleveref}
\usepackage[tight,footnotesize]{subfigure}
\begin{document}
%
\title{Nonparametric Identification and Estimation of Earnings Dynamics using a Hidden Markov Model: Evidence from the PSID}


\author{\IEEEauthorblockN{Tong Zhou}
  \IEEEauthorblockA{Department of Computer Science\\
Johns Hopkins University\\
Baltimore, United States\\
Email: tzhou11@jhu.edu\\
\textit{DOI: } \href{https://doi.org/\mydoi}{\mydoi}
}
}


%


\maketitle

\thispagestyle{fancy}
\lhead{}
\lfoot{}
\cfoot{\small{© © 2023 IEEE. Personal use of this material is permitted. Permission from IEEE must be obtained for all other uses, in any current or future media, including reprinting/republishing this material for advertising or promotional purposes, creating new collective works, for resale or redistribution to servers or lists, or reuse of any copyrighted component of this work in other works.}}
\rfoot{}

\begin{abstract}
This paper presents a hidden Markov model designed to investigate the complex nature of earnings persistence. The proposed model assumes that the residuals of log-earnings consist of a persistent component and a transitory component, both following general Markov processes. Nonparametric identification is achieved through spectral decomposition of linear operators, and a modified stochastic EM algorithm is introduced for model estimation. Applying the framework to the Panel Study of Income Dynamics (PSID) dataset, we find that the earnings process displays nonlinear persistence, conditional skewness, and conditional kurtosis. Additionally, the transitory component is found to possess non-Gaussian properties, resulting in a significantly asymmetric distributional impact when high-earning households face negative shocks or low-earning households encounter positive shocks. Our empirical findings also reveal the presence of ARCH effects in earnings at horizons ranging from 2 to 8 years, further highlighting the complex dynamics of earnings persistence.
  \end{abstract}


%
\IEEEpeerreviewmaketitle

\section{Introduction}
Earnings dynamics is a fascinating and important area in economics, with significant implications for understanding economic agents' consumption decisions. Macroeconomists employ life-cycle models and profiles of agents' earnings dynamics to examine their various responses within the economy, laying the foundation for the creation of sensible policies to manage business cycles. In a broader context, the nature of earnings dynamics is crucial in addressing a wide range of economic issues, including income inequality, optimal design of fiscal policies and insurance programs, economic mobility, and human capital development. As such, accurately characterizing earnings dynamics enables more effective management and a deeper understanding of a country's economy.

We utilize a parsimonious specification of the earnings process, where log-earnings consist of an unobserved persistent shock and an unobserved transitory shock. The literature on earnings process specifications varies in its focus on the distinction between these two types of shocks, a concept that can be traced back to Nobel laureate Milton Friedman's renowned permanent income hypothesis (PIH). Although there are numerous models of earnings dynamics, most tend to concentrate on linear specifications for these two hidden components, inherently excluding the possibility of nonlinear transmission of earnings shocks.

In this paper, we introduce a new nonparametric framework to explore earnings dynamics. Both the persistent and transitory components are modeled as two generic first-order Markov processes. Apart from the first-order restriction, no further assumptions are imposed on the model. In essence, our specification establishes a hidden Markov model (HMM) with two latent state variables. Our focus is on identifying the two Markov kernels, specifically, the conditional distributions of the persistent component given its past and the conditional distribution of the transitory component given its past.

We propose a two-step stochastic EM algorithm for estimating the model. In the E-step, we use an MCMC procedure to obtain draws for the two hidden components through a likelihood-based approach. In the M-step, we perform a maximization procedure on a series of quantile regressions with imputed values for hidden covariates. The iteration continues until the expected likelihood is maximized.
\section{Materials and Methods}

\subsection{Model}
\subsubsection{Setup}
In line with the conventions of earnings dynamic literature, we use $\log Y$ to represent the real (log) earnings, and it can be decomposed into the explanatory part, a persistent component $U$ and a transitory component  $V$. The earnings process for each household  $i$ at time  $t$ is as follows:
\begin{equation}
  \log(Y_{it}) = \mathbf{z}_{it}^\prime \bm{\beta}  + U_{it} + V_{it}, 
\end{equation}
where $\mathbf{z}_{it} $ is a set of observed demographics and known by agents at $t$. We let  $y_{it} = \log(Y_{it}) - \mathbf{z}_{it}^\prime \bm{\beta}  $ denote the log of real income net of predictable individual components.

We assume both $U_{it}$  and $V_{it}$ follow some general unknown functions $H_{t}(U_{i, t - 1}, \eta_{it})$ and $Q_{t}(V_{i, t-1}, \varepsilon_{it})$, where $\eta_{i t}$ and $\varepsilon_{it}$ are assumed to follow conditional standard uniform distributions, i.e. 
\begin{eqnarray}
  \eta_{it} | (U_{i, t - 1}, U_{i, t - 2}, \cdots) &=  \mathsf{Unif}(0, 1), ~ ~ t = 2, \cdots, T \\  
  \varepsilon_{it} | (V_{i, t - 1}, V_{i, t - 2}, \cdots) &=  \mathsf{Unif}(0, 1), ~ ~ t = 2, \cdots, T .
\end{eqnarray}
This general nonparametric setting offers greater flexibility for studying the persistence of earnings dynamics and encompasses many earnings dynamic models as special cases including the \textit{canonical earnings dynamics models}, where the persistent component follows a unit-root process. Given that both processes are unobserved, a Bernoulli instrumental variable $\omega(C_{it})$ is required to differentiate them, where $\omega(\cdot)$ is a known transformation of consumption data $C_{it}$ for agent $i$ at  $t$. Since the purpose of $\omega(C_{it})$ is merely to distinguish the two Markov kernels, it suffices for our purposes to use a logistic function, i.e. $\mathbb{P}(\omega(C_{it}) = 1 | U_{it}) = 1 / (1 + \exp(-\beta_0 - \beta_1 U_{it} ))$. The rationale behind this setup can be found in the works of \cite{blundell2008, blundell2016, blundell2018}.

Putting above discussions together, we have the complete model setup
\begin{eqnarray*}
 y_{it} &=&  U_{it} + V_{it} \\ 
 U_{it} &=&  H_{t}(U_{i, t-1}, \eta_{it}) \\
 V_{it} &=&  Q_{t}(V_{i, t-1}, \varepsilon_{it}) \\
 \mathbb{P}(\omega(C_{it}) = 1) &=&  1 / (1 + \exp(-\beta_0 - \beta_1 U_{it} ))
\end{eqnarray*}

\subsubsection{Assumptions}
We will outline the assumptions needed to identify the model. Our identification strategy relies on the powerful spectral decomposition of linear operators. A thorough overview of this approach can be found in the work of \cite{Schennach}.
\begin{assumption}
  \begin{enumerate}
    \item (First-order Markov) Both $U_{it}$ and $V_{it}$ follow a generic first-order Markov process;
    \item (Conditional uniform distribution) Both $\eta_{it}$ and $\varepsilon_{it}$ follow conditional standard uniform distributions 
    \item (Monotonicity) The unknown condition quantile function $\tau \mapsto H_{t}(U_{i, t-1}, \tau)$ and $\tau \mapsto Q_{t}(V_{i, t-1}, \tau)$ are strictly increasing for $\tau \in (0, 1)$. 
    \item (Invertibility) The conditional distribution functions $F(U_{it} | U_{i, t-1})$ and $F(V_{it} | V_{i, t-1})$ are both invertible w.r.t. their respective arguments $U_{it}$ and $V_{it}$ for each $i$ and  $t$.
  \end{enumerate}
\end{assumption}
Assumption 1) states that $U_{it}$ and $V_{it}$ have only one-period memory of their past. This condition imposes dynamic exclusion restrictions that aid in obtaining nonparametric identifications. This assumption is also commonly made in structural economic models. Although it can be relaxed to allow for higher-order Markov process, we maintain the first-order Markovian assumption in this paper for simplicity. Assumption 2) normalizes the error terms $\eta_{it}$ and $\varepsilon_{it}$ to follow standard uniform distributions. This setup enables us to discuss consequences of shocks along the rank of $U_{i, t-1}$ and $V_{i, t-1}$. This representation also nests the canonical model of earnings dynamics as a special case where $U_{it}$ is assumed to follow a unit-root process, i.e. 
\begin{equation*}
  U_{i, t+ 1} = U_{it} + \nu_{i, t+ 1} 
\end{equation*}
where $\nu_{i, t+1} = F^{-1}(\eta_{i, t+1})$ is the inverse function of the CDF of $\eta_{i, t+1}$. Assumption 3) guarantees that $U_{it}$ and $V_{it}$ have absolutely continuous distributions. Assumption 1) - 3) combined imply that for all $\tau \in (0,1)$, $H_{t}(U_{i,t-1}, \tau)$ happens to be the $\tau$-conditional quantile of $U_{it}$ given $U_{i, t-1}$. This relationship also holds for $Q_{t}(V_{i, t-1}, \tau)$. Assumption 4) is furnished to facilitate identification of the nonlinear functions $H_{t}$ and $Q_{t}$. The monotonicity restriction on $H_{t}$ and $Q_{t}$ are necessary for the existence of their marginal densities $f(U_{it} | U_{i, t-1})$ and $f(V_{it} | V_{i, t-1})$. It is not a sufficient condition because a stronger condition of absolute continuity on the distribution function $F_{V_{t}|V_{t-1}}$  cannot be weakened. However, since it is rare that a distribution function is continuous but not absolutely continuous, assumption 4) can be almost equivalent to the existence of the two marginal densities.

\begin{assumption}[independence]
  Two random vectors $(\eta_{i 2}, \cdots, \eta_{i T}, U_{i 1})$ and $(\varepsilon_{i 2}, \cdots, \varepsilon_{i T}, V_{i 1})$ are statistically independent. The Bernoulli random variable $\omega(C_t)$ is independent of $V_t$ for all $t$.
\end{assumption}
This assumption suggests that the persistent process $\left\{ U_{it}  \right\}_{t=1}^{T} $ and the transitory process $\left\{ V_{it}  \right\}_{t=1}^{T} $ are statistically independent. This restriction allows for the common deconvolution technique of separating two unknown probability densities. For instance, once one of the marginal densities $f(U_{it})$ or $f(V_{it})$ is identified, the other one will also be automatically identified through the deconvolution argument.

Since our identification strategy relies on the technique of manipulating linear operators, we provide the definition of linear operator here to facilitate our later discussions. Let $\mathcal{L}^{p}(F_{U})$ denote the collection of functions of variable $U$ for which its  $p$-th moment is finite, i.e.  $g \in \mathcal{L}^{p}(F_{u})$ implies
\begin{equation}
  \|g\|_{\mathcal{L}^{p}} = \left( \int_{\mathcal{U}} g(u) \mathrm{d}F_{U}(u)  \right)^{\frac{1}{p}} < \infty, 
\end{equation}
where $\mathcal{U}$ denotes the support of $U$. The definition for the space $\mathcal{L}^{q}(F_{V})$ is similar.

Now we define a linear operator
\begin{equation}
  \mathcal{L}_{V | U}: \mathcal{L}^{p}(F_{U})    \to \mathcal{L}^{q}(F_{V}), 
\end{equation}
where $p, q \geq 1$. Specifically, for any $g \in \mathcal{L}^{p}(F_{U})$, we have
\begin{equation}
  \mathcal{L}_{V|U}g = \int_{\mathcal{U}} f_{V|U}(v|u) g(u) \mathrm{d}u \in \mathcal{L}^{q}(F_{V}), 
\end{equation}
where the function $f_{V|U}$ is called the kernel of the linear operator $\mathcal{L}_{V|U}$. This expression is particularly useful when multiple linear operators are present, since we do not need to introduce new notations for each involved operator.

\begin{assumption}
  There exist variables $Y_{it}$ such that 
  \begin{enumerate}
    \item For any $y_{t}$ can $\widetilde{c}_{t} $, there exists a $y_{ t-1}$ and $\widetilde{c}_{t-2} $ and a neighborhood $\mathcal{N}^{r}$ around $(y_{t}, \widetilde{c}_{t-1}, y_{t-1}, \widetilde{c}_{t-2}  )$ such that, for any $(y_{t}^\prime, \tilde{c}_{t-1}^\prime, y_{t-1}^\prime, \tilde{c}_{ t-2}^\prime) \in \mathcal{N}^{r}$, the linear operator $\mathcal{L}_{Y_{t-2}, y_{t-1}^\prime, \tilde{c}_{t-2}^\prime, y_{t}^\prime, \tilde{c}_{t-1}^\prime, Y_{t+1}}$ is one-to-one.
    \item For nay $y_{t}$ and $\tilde{c}_{t-1}$, the linear operator $\mathcal{L}_{Y_{t=1}|y_{t}, \tilde{c}_{t-1}, U_{t-1}, V_{t}}$ is one-to-one.
    \item For any $y_{t-1}$ and $\tilde{c}_{t-2}$, the linear operator $\mathcal{L}_{Y_{t-2}, y_{t-1}, \tilde{c}_{t-2}, Y_{t}}$ is one-to-one.
  \end{enumerate}
\end{assumption}

\begin{assumption}
  \begin{enumerate}
    \item The characteristic function of $(U_{i 1}, \cdots, U_{i T})$ and $(V_{i 1}, \cdots, V_{i T})$ do not vanish on the real line.
    \item The characteristic function of $(U_{i 1}, \cdots, U_{i T})$ and $(V_{i 1}, \cdots, V_{iT})$ are absolutely continuous.
  \end{enumerate}
\end{assumption}
Assumption 4.1) is commonly made to achieve nonparametric identification (see \cite{haul2011}). For univariate distributions, this assumption rules out certain families of distributions, e.g., truncated normal, symmetric uniform and many discrete distributions. Assumption 4.2) is made to facilitate the deconvolution argument and also implies that the joint distributions of $(U_{i 1}, \cdots, U_{iT})$ and $(V_{i 1}, \cdots, V_{iT})$ exist.

To avoid cluttered notations, we simplify the notations by omitting the subscript $i$ without causing confusions. In the following derivations, we define  $\widetilde{C}_{t-1} :=  \omega(C_{t-1})$.
\begin{assumption}[Uniqueness of spectral decomposition]
For any $(Y_{t}, \widetilde{C}_{t - 1} )$ and any $(u_{t - 1},v_{t}) \neq (u_{t-1}^\prime, v_{t}^\prime)$, there exists a $(y_{t-1}, \widetilde{c}_{t-2} )$ and corresponding neighborhood $\mathcal{N}^{r}$ satisfying Assumption 3.1), such that for some $(y_{t}^\prime, \tilde{c}_{t-1}^\prime, y_{t-1}^\prime, \tilde{c}_{t-2}) \in \mathcal{N}^{r}$ with $(y_{t}^\prime, \tilde{c}_{t-1}) \neq (y_{t}, \tilde{c}_{t-1})$ and $(y_{t-1}^\prime, \tilde{c}_{t-2}) \neq (y_{t-1}, \tilde{c}_{t-2})$: 
    \begin{equation*}
      0 < k(y_{t}, \tilde{c}_{t-1}, y_{t-1}^\prime, \tilde{c}_{t-2}, y_{t-1}, \tilde{c}_{t-2}, u_{t-1}, v_{t}) < C < \infty
    \end{equation*}
    and
    \begin{eqnarray*}
      k(y_{t}, \tilde{c}_{t-1}, y_{t-1}^\prime, \tilde{c}_{t-2}, y_{t-1}, \tilde{c}_{t-2}, u_{t-1}, v_{t}) \neq  \\ k(y_{t}, \tilde{c}_{t-1}, y_{t-1}^\prime, \tilde{c}_{t-2}, y_{t-1}, \tilde{c}_{t-2}, u_{t-1}^\prime, v_{t}^\prime)  
    \end{eqnarray*}
 where 
 \begin{align*}
   & k(y_{t}, \tilde{c}_{t-1}, y_{t-1}^\prime, \tilde{c}_{t-2}, y_{t-1}, \tilde{c}_{t-2}, u_{t-1}, v_{t}) = \\
 &\frac{f(y_{t}, \tilde{c}_{t-1}|y_{t-1}, \tilde{c}_{t-2}, u_{t-1}, v_{t} )f(y_{t}^\prime, \tilde{c}_{t-1}^\prime|y_{t-1}^\prime, \tilde{c}_{t-2}^\prime, u_{t-1}, v_{t}))}{ f(y_{t}^\prime, \tilde{c}_{t-1}^\prime|y_{t-1}, \tilde{c}_{t-2}, u_{t-1}, v_{t} f(u_{t-1}, v_{t}|y_{t-1}^\prime, \tilde{c}_{t-2}^\prime, u_{t-1}, v_{t}))}
 \end{align*}
\end{assumption}
\begin{assumption}[normalization]
  The Markov kernels are normalized by $\mathbb{E}[U_{t+1}|U_{t}] = U_{t}$ and $\mathbb{E}[V_{t+1}|V_{t}] = 0$.
\end{assumption}
In the eigenfunctions $f(y_{t+1} | y_{t}, \tilde{c}_{t-1}, u_{t-1}, v_{t})$, both $u_{t-1}$ and $v_{t}$ are unobserved and continuously distributed. Assumption 6 is made to differentiate and identify the two  components. 

\begin{assumption}[Stationarity]
  For any $2 \le  t \le  T$, the Markov kernels is time-invariant, i.e., 
  \begin{eqnarray*}
    f(Y_{t}, \widetilde{C}_{t-1}, U_{t-1}, V_{t} | Y_{t-1}, \widetilde{C}_{t-2}, U_{t-2}, V_{t-1}  ) \\ 
    = f(Y_{3}, \widetilde{C}_{2}, U_2, V_3 | Y_2, \widetilde{C}_{1}, U_1, V_2  )
  \end{eqnarray*}
\end{assumption}
Assumption 7 is not necessary for identification of the Markov density. It eases our derivations. From the next section, we can see that only five periods of data is sufficient for achieving nonparametric identification.

\subsection{Identification}

\begin{figure}[h]
  \begin{center}
    \begin{tikzpicture}[->, >=stealth',thick,node distance=2.5cm,scale=0.4]
    \tikzstyle{every state} = [draw=none,very thick,fill=blue!20]
    \tikzstyle{epsilon}=[green!50!black,text=white,draw=none,very thick=blue!20]
    \tikzstyle{consumption}=[red,text=white,draw=none]
    \tikzstyle{yy} = [rectangle]
    \tikzstyle{uu} = [orange,text=white]
    \node[state,yy]     (Y2)     {$(Y_{t-2},\widetilde{C}_{t-3}) $};
    \node[state,yy]     (Y1)[right of =Y2]      {$(Y_{t-1},\widetilde{C}_{t-2}) $}  ;
    \node[state,yy]     (Y0)[right of =Y1]      {$(Y_{t},\widetilde{C}_{t-1}) $};
    \node[state,yy]     (Y11)[right of =Y0]      {$(Y_{t+1},\widetilde{C}_{t}) $};
    \node[state,uu]     (U2)[below of =Y2]       {$U_{t-2}$} ;
    \node[state,uu]     (U1)[below of =Y1]       {$U_{t-1}$} ;
    \node[state,uu]     (U0)[below of =Y0]       {$~U_{t~~}$} ;
    \node[state,uu]     (U11)[below of =Y11]       {$U_{t+1}$} ;
    \node[state,epsilon]      (E2)[above of = Y2]       {$V_{t-2}$};
    \node[state,epsilon]      (E1)[above of = Y1]       {$V_{t-1}$};
    \node[state,epsilon]      (E0)[above of = Y0]       {$~V_{t~~}$};
    \node[state,epsilon]      (E11)[above of = Y11]       {$V_{t+1}$};
    \path[->]
    (U2)              edge[right]              (U1)
                      edge[above]              (Y2)
                      edge                      (Y1)
    (U1)              edge[right]              (U0)
                        edge[above]              (Y1)
                        edge                    (Y0)
    (U0)              edge[right]              (U11)
                        edge[above]              (Y0)
                        edge                   (Y11)
    (U11)       edge[above]              (Y11)
    (E2)              edge[below]              (Y2)
                      edge[right]              (E1)
    (E1)              edge[below]              (Y1)
                      edge[right]              (E0)
    (E0)              edge[right]              (E11)
                      edge[below]              (Y0)
    (E11)              edge[below]              (Y11)
    (Y2)              edge[right]              (Y1)
    (Y1)              edge[right]              (Y0)
    (Y0)              edge[right]              (Y11)
    (-3,-3)            edge                   (Y2);
\end{tikzpicture} 
  \end{center}
  \caption{Graphical illustration of earnings dynamics }
  \label{fig:flow}
\end{figure}
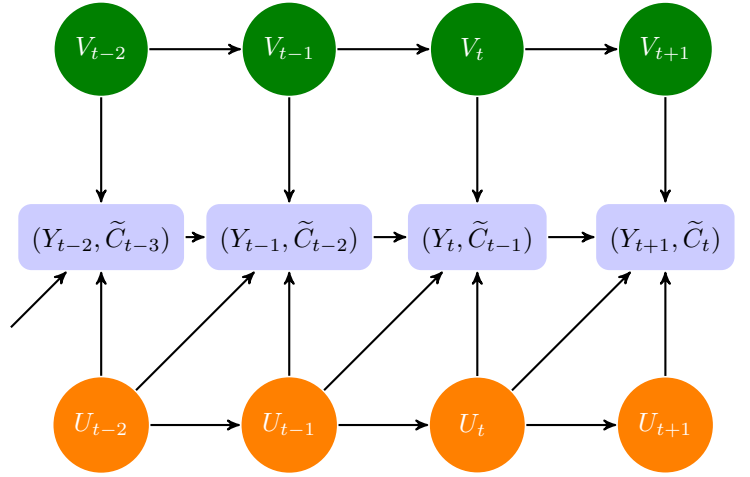

Based on the assumptions made in the previous section, identification can be accomplished by applying Theorem 9 from \cite{Schennach} in dynamic settings. This strategy can be better understood in \Cref{fig:flow}, where the dependence structures and dynamic exclusion restrictions can be easily visualized.

We state the main identification theorem
\begin{theorem}[Identification]
  \label{thm}
  Under Assumption 1 - Assumption 7, the density 
    $$f(Y_{t+1}, \widetilde{C}_{t}, Y_{t}, \widetilde{C}_{t-1}, Y_{t-1}, \widetilde{C}_{t-2}, Y_{t-2}, \widetilde{C}_{t-3}    )$$
  for any $t \in \left\{ 4, \cdots, T-1  \right\} $ uniquely determines the densities $f(Y_{t}, \widetilde{C}_{t-1}, U_{t-1}, V_{t}| Y_{t-1}, \widetilde{C}_{t-2}, U_{t-2}, V_{t-1}  )$.
\end{theorem}
\Cref{thm} implies that our interests of Markov kernels can be identified by basic probability rules, the Bayes rules and the deconvolution technique.
\begin{corollary}
  Under Assumption 1 - Assumption 7, the Markov kernels $f_{V_{t}|V_{t-1}}$, $f_{U_{t}|U_{t-1}}$ and marginal distributions $f_{U_{t}}$ and $f_{V_{t}}$ are uniquely identified, for $t = 4, \dots, T-1$.
\end{corollary}

\subsection{Estimation}
We introduce a \textit{modified stochastic EM algorithm} (MSEM) to estimate this HMM,  while the stochastic EM was originally proposed by \cite{Nielsen2000}. The MSEM provides a much more faster implementation of the estimation by replacing the likelihood with the objective functions of quantile regression models. The MSEM is similar to the one presented in \cite{arellano2017}. The difference lies in the fact that their paper involves only one state variables. Specifically, for any $\tau \in (0, 1)$ we employ the following estimating equations
\begin{align*}
  U_{it} &= \sum_{k =0}^{K_1}a_{k}^{H}(\tau) \phi_{k}(U_{i, t-1}, \mathrm{age}_{it}) \\
  V_{it} &= \sum_{k=0}^{K_2} a_{k}^{Q}(\tau) \phi_{k}(V_{i, t-1}, \mathrm{age}_{it}) \\
  U_{i 1} &=\sum_{k=0}^{K_3}a_{k}^{H_1}(\tau) \phi_{k}(\mathrm{age}_{i 1}) \\
  V_{i 1} &=  \sum_{k=0}^{K_2i}a_{k}^{Q_1}(\tau) \phi_{k}(\mathrm{age}_{i 1}), 
\end{align*}
where $\phi_{k}$ is the Hermite polynomials. We selected different orders of polynomials for the four equations to maximize the likelihood. The quantile-based estimation strategy provides a flexible specification of the Markov kernels. \cite{arellano2016} applies this estimation strategy to estimate the smoking effects of women during pregnancy on children's birthweights.

Another advantage of using quantile-based estimation is that the original nonparametric estimation problem is reduced to  estimating a finite number of parameters, i.e., the coefficients of the Hermite polynomials. We discuss $U_{it}$ as an example: the functions $a_{k}^{H}(\tau)$ are modeled as piecewise-polynomial interpolating splines on equi-length intervals $[\tau_1, \tau_2], [\tau_2, \tau_3], \cdots, [\tau_{I - 1}, \tau_{I}]$ that partition the unit interval $(0,1)$. In other words, we need to estimate  $a_{k}^{H}(\tau)$ for each interval of $\tau$ and $k$. Additionally, the objective function of quantile regressions can be used as a surrogate likelihood. Since it is a convex function, the implementation can be fast.  Once those  $a_{\tau}^{H}$ 's are obtained, we are finished with the estimation of $U_{it}$.

We still take $U_{it}$ as an example to illustrate the MSEM algorithm. We start with an initial value for the parameter vector  $\widehat{\theta}^{(0)} $. Each iteration follows the following two steps until convergence of the $\widehat{\theta}^{(s)} $ in the $s$-th iteration:
  
  \begin{itemize}
    \item \emph{Stochastic E-step:} Draw $U_{i}^{(m)} = (U_{i 1}^{(m)}, \cdots, U_{i T}^{(m)} )$ for $m = 1,\cdots, M $ from $f_{i}(\cdot ; \widehat{\theta}^{(s)}  )$.
    \item \emph{M-step:} Compute 
      \[
        \widehat{\theta}^{(s+1)} =    \operatorname*{arg\,min}_{\theta} \sum_{i=1}^{N}\sum_{m=1}^{M}R(y_{i}, U_{i}^{(m)}; \theta),
      \]
  \end{itemize}
where $R(\cdot)$ is the surrogate likelihood, i.e. the objective function of the piece-wise quantile regressions. In the E-step, we use a random-walk Metropolis-Hastings algorithm for drawing $U_{i}^{(m)}$ in the E-step. The M-step consists of a number of quantile regressions. For instance, for each $\ell$,  the parameters $a_{k}^{H}(\tau_{\ell})$ are updated as 
  \begin{align*}
  \min_{(a_{0 \ell}^{H},\cdots, a_{K \ell}^{H} )} \sum_{i=1}^{N}&\sum_{t=2}^{T}\sum_{m=1}^{M}  \\  & \rho_{\tau_{\ell}}\left(  U_{it}^{(m)} - \sum_{k=0}^{K}a_{k \ell}^{H} \varphi_{k}\left( U_{i,t-1}^{(m)}, \textsf{age}_{it}\right) \right),
  \end{align*}
  where $\rho_{\tau}(u) = u (\tau - \bm{1}(u \leq 0))$ is the check function in standard quantile regressions and $\bm{1}(\cdot)$ denotes the indicator function, first introduced by \cite{koenker1978}. 

\begin{figure}[h!]
  \centering
 \includegraphics[width=.5\textwidth]{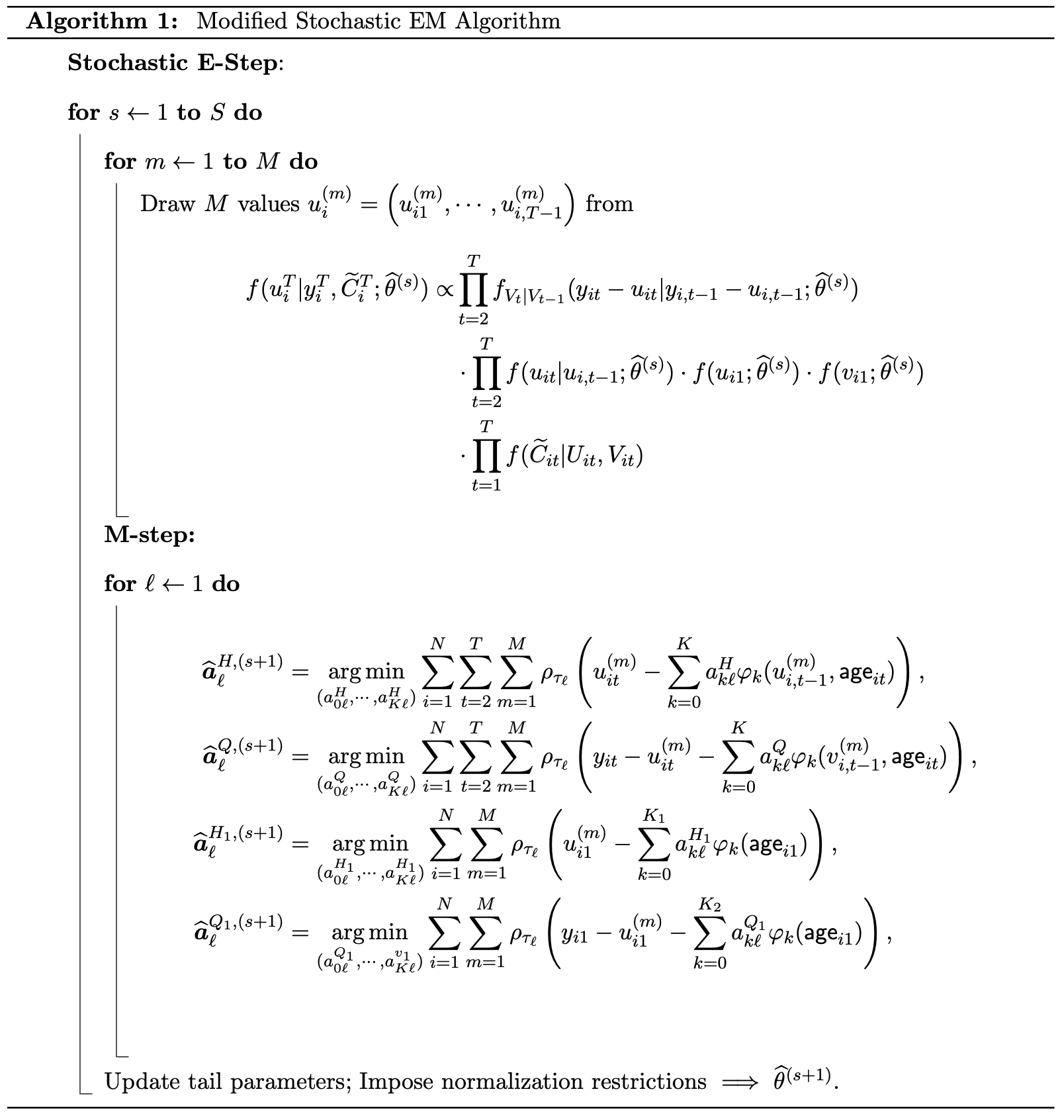}
\end{figure}
\cite{Nielsen2000} examined the statistical properties of the stochastic EM algorithm within a likelihood case. He provided certain conditions under which the Markov chain $\widehat{\theta}^{(s)} $ is ergodic. He also outlined the asymptotic distribution of $\widehat{\theta} $. \cite{arellano2017}  characterized the asymptotic distribution of $\widehat{\theta} $ in a manner that aligns with our model, specifically when utilizing the surrogate likelihood during the M-step.

The MSEM algorithm can be summarized as follows

  \section{Results and Discussions}
  \subsection{Data}
  As the longest running household panel survey in the world, the PSID dataset contains a large amount of data in the US over 50 years. The dataset includes a wide range of variables on income, employment, education, economic, social, health-related factors and many other aspects of life for each individual and their family members. However, the size, variety and complexity of the dataset make it challenging to analyze using traditional statistical methods.

  The second challenge comes from the longitudinal nature, which means that it follows the same individuals and families over time. This presents unique challenges for analysis, such as handling missing data, attrition, and changes in the variables of interest over time.

  The third one is data quality. The quality of PSID can vary over time, as changes in survey methodology or sample composition can affect the accuracy and reliability of the data. This requires careful attention to data cleaning and quality control procedures.

  Our study draws on the PSID as our primary data source. To ensure high quality data and better compare our empirical findings with other literature, our sample selection procedure and data preprocessing mainly follow the works of \cite{blundell2008, blundell2016}.

  \subsection{Empirical Findings}
  \subsubsection{Densities and Moments}
\begin{figure}[h]
	\subfigure{       
      \begin{minipage}{2.5cm}                                                          \includegraphics[scale=0.14]{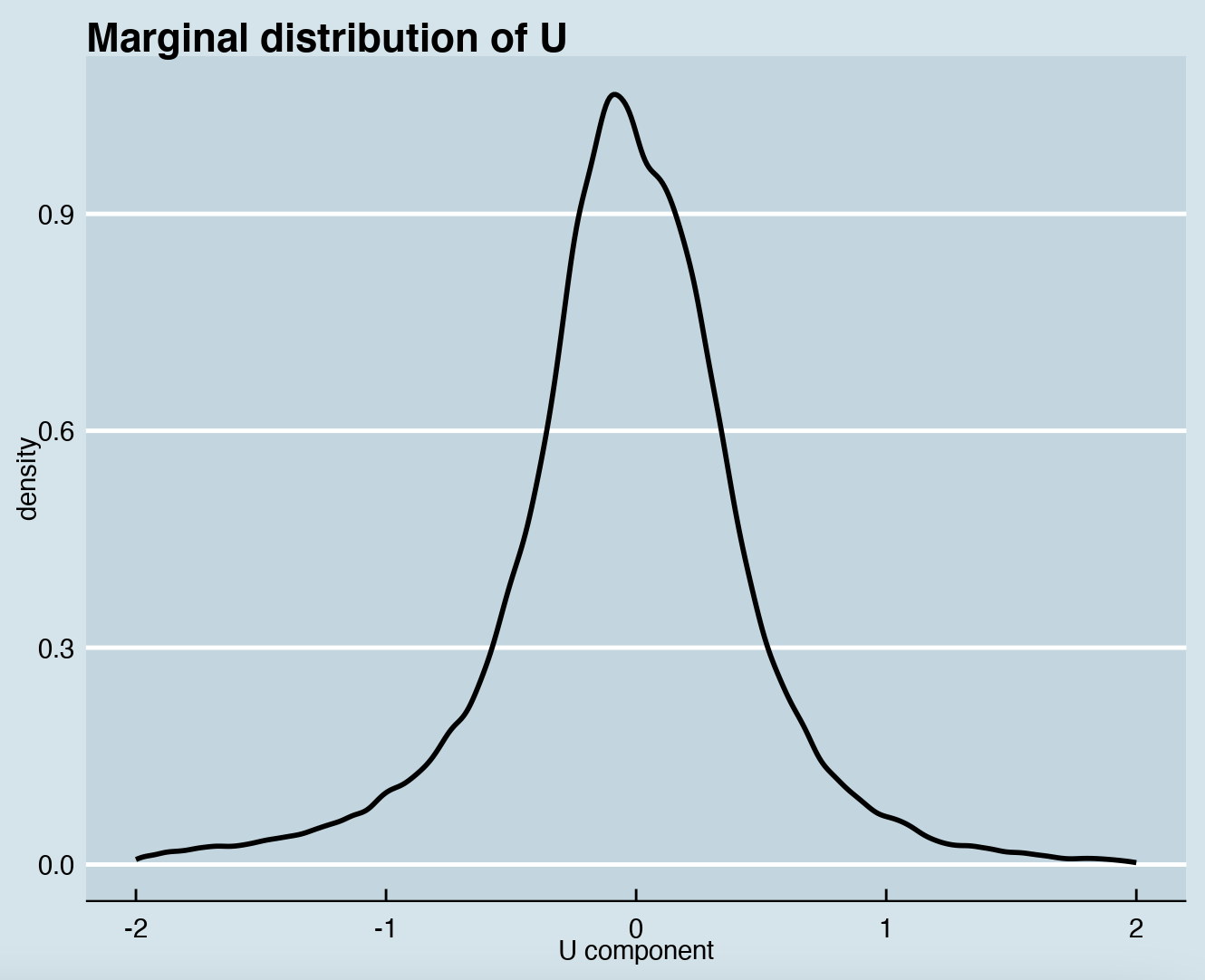} 
	\end{minipage}} 
    \hspace{4em}
	\subfigure{  
      \begin{minipage}{2.5cm}                                                          \includegraphics[scale=0.14]{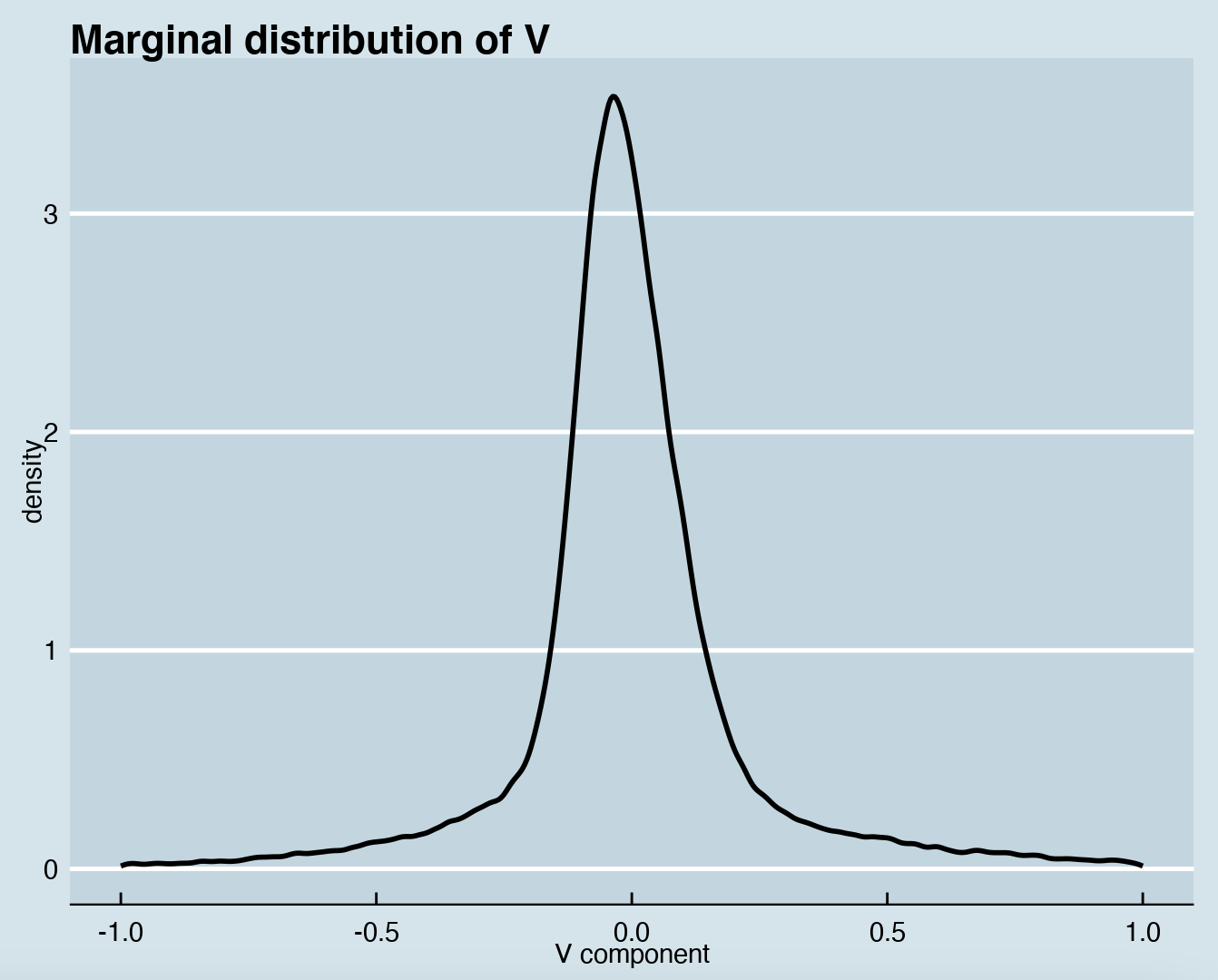}  
	\end{minipage}}
	\caption{Marginal distributions of persistent and transitory earnings components} 
	\label{fig:marginal}  
\end{figure}

\begin{figure}[h]
	\subfigure{       
      \begin{minipage}{2.5cm}                                                          \includegraphics[scale=0.14]{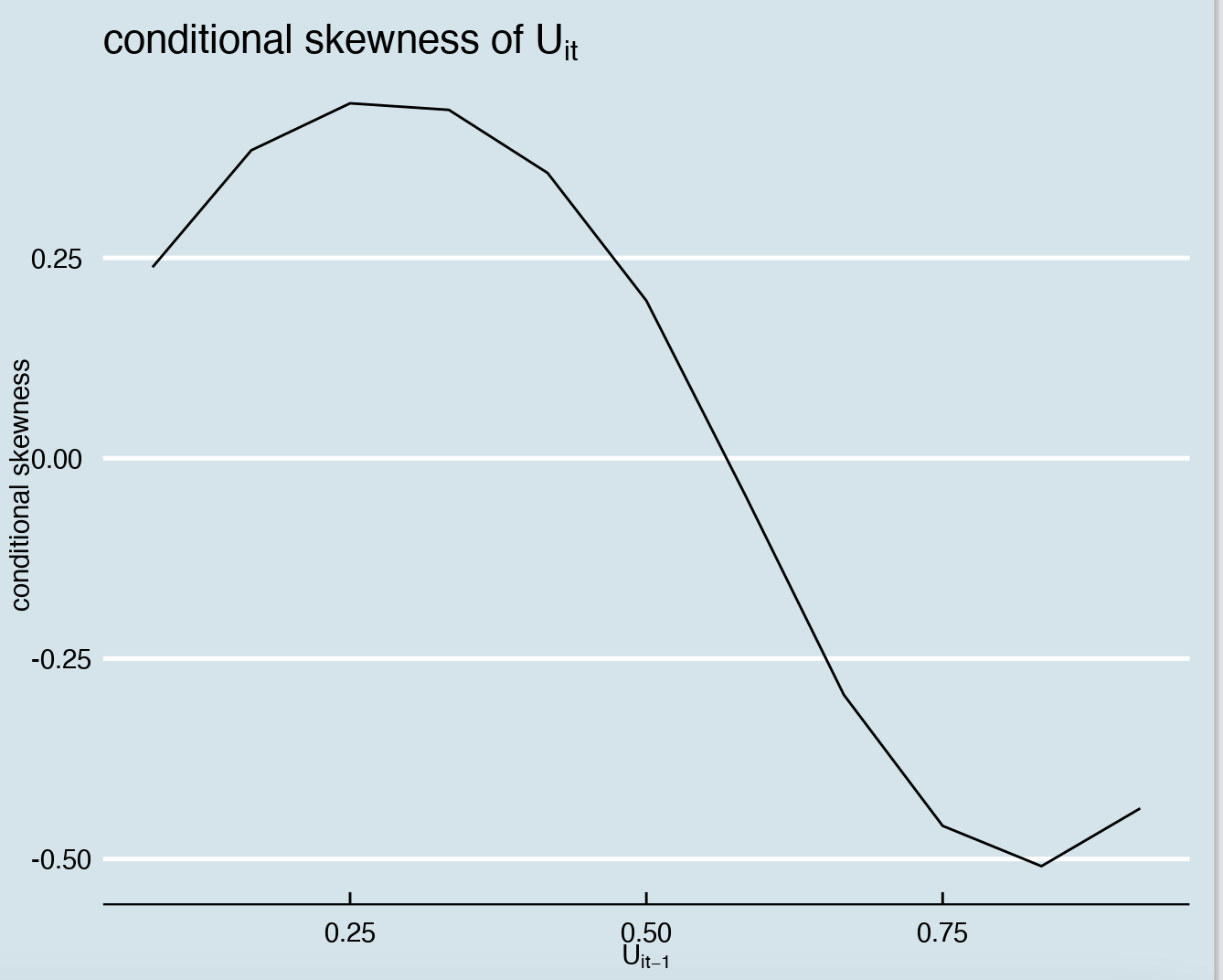}
	\end{minipage}} 
    \hspace{4em}
	\subfigure{  
      \begin{minipage}{2.5cm}                                                          \includegraphics[scale=0.14]{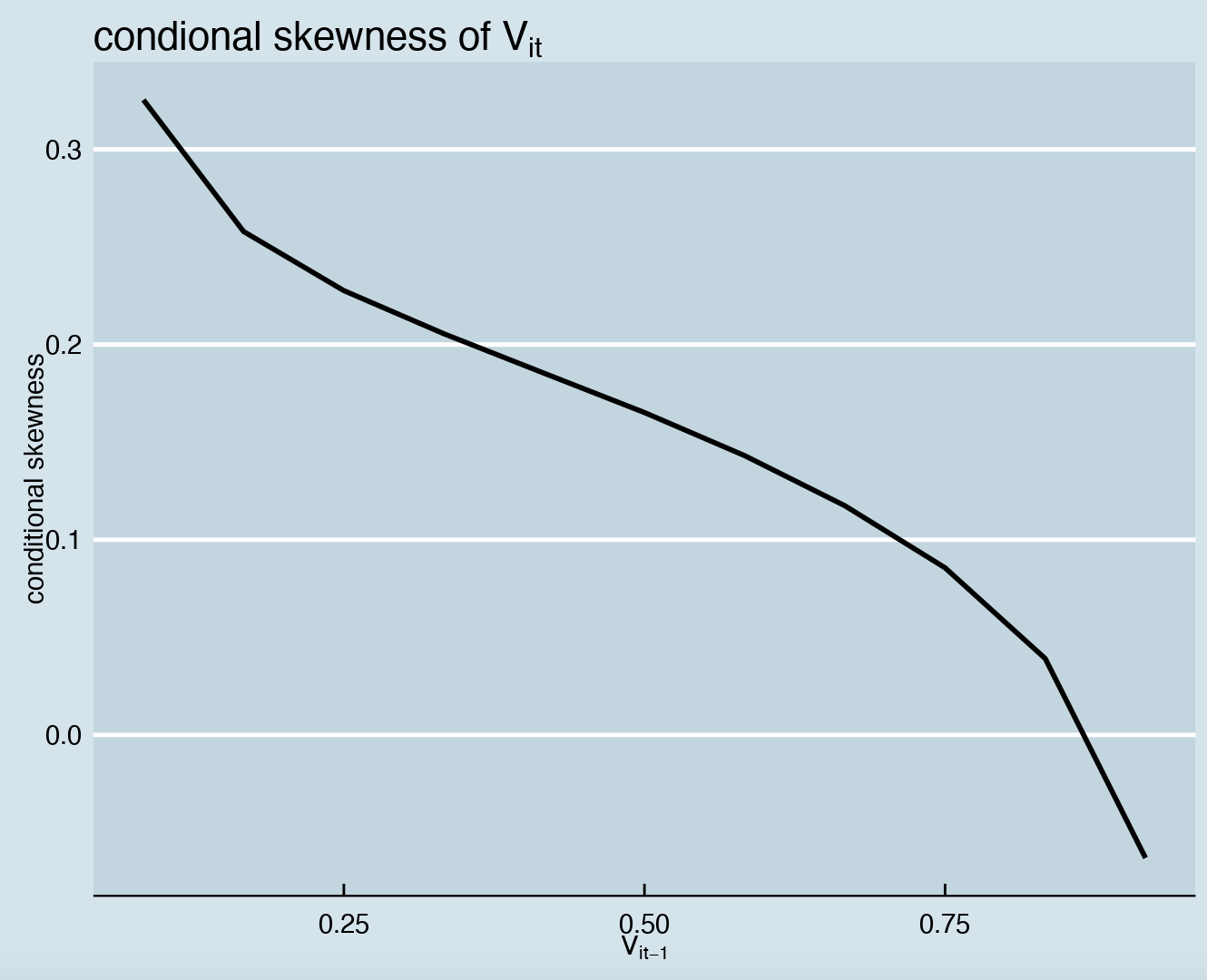}  
	\end{minipage}}
	\caption{Conditional skewness of $U$ component and $V$ component.} 
	\label{fig:skew}  
	\end{figure}

  \Cref{fig:marginal} illustrates the marginal distributions of the persistent and transitory earnings components at the mean age. The persistent component $U_{it}$ displays small deviations from Gaussianity. However, the marginal distribution of $V_{it}$ provides strong evidence to reject Gaussianity owing to its high kurtosis and fat tails. It is worth noting that the density of $V_{it}$ in our model is less spiky than that in \cite{arellano2017}. A possible explanation for this difference could be the mutual dependence structure of $V_{it}$, which is governed by its first-order Markovian property, whereas they are assumed to be mutually independent across $t$ in their paper. In \Cref{fig:skew}, we report the conditional sknewness for $\tau = 11/ 12$, for both the $U$ component and the $V$ component. These two panels show a similar pattern: $U_{it}$($V_{it}$) is positively skewed for low values of $U_{i,t-1}$($V_{i,t-1}$), and negatively skewed for high values of $U_{i,t-1}$($V_{i,t-1}$). 

\subsubsection{Nonlinear persistence}
\begin{figure}[h!]
 \centering 
 \includegraphics[width=.36\textwidth]{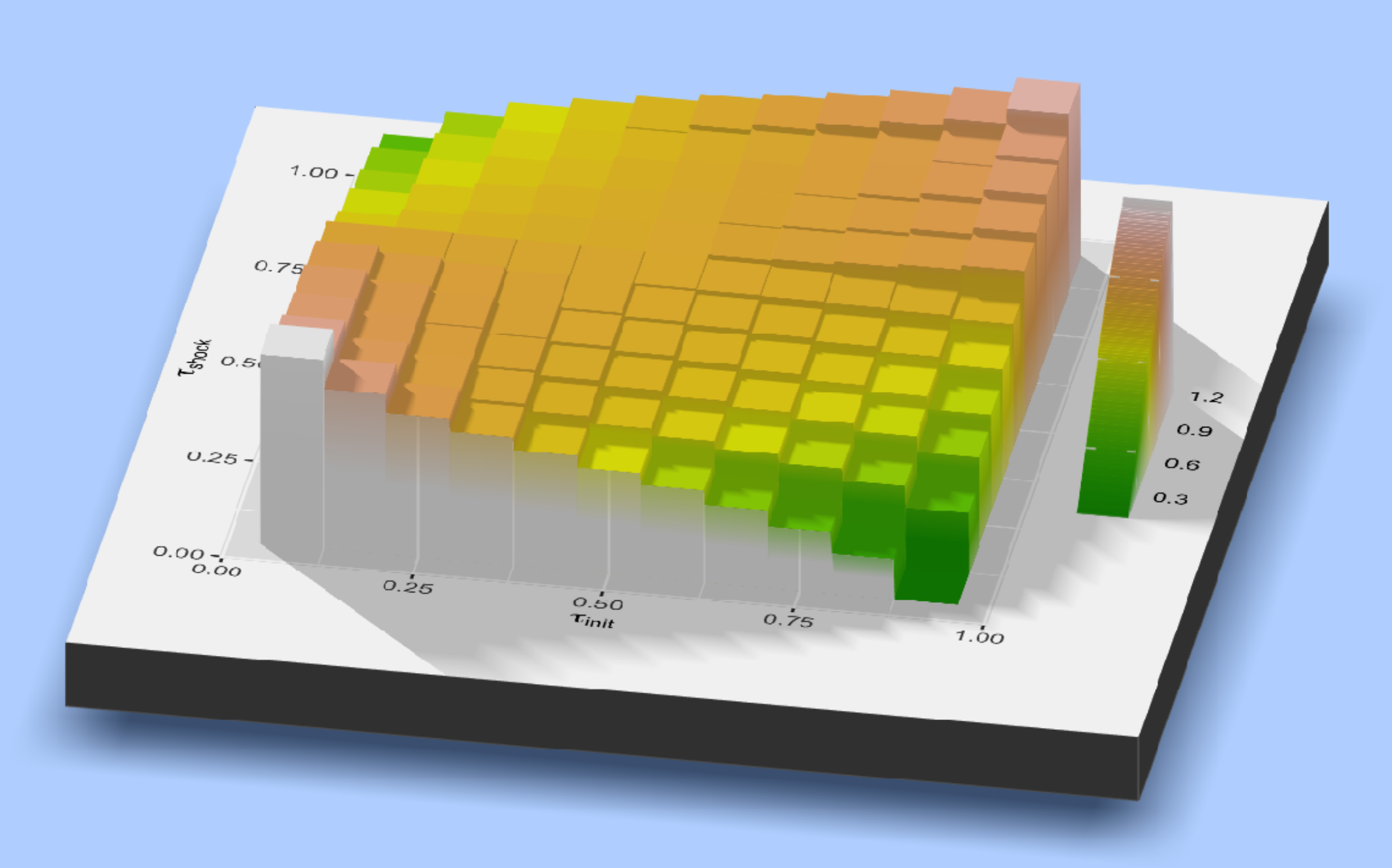}
 \caption{Estimates of the average derivative of the conditional quantile function of simulated model.}
 \label{fig:simulatedY}
\end{figure}

This experiment examines the marginal effects of persistent and transitory shocks. For linear specifications of earnings dynamics, their marginal effects would be constant by design. In the canonical model of earnings dynamics, for example, where the innovation is a random walk, then the marginal effect is $1$, regardless of $U_{i,t-1}$ and $\tau$. In contrast, our model allows the persistence of $U_{i,t-1}$ to depend on the magnitude and direction of the shock. As a result, the persistence of a shock to $U_{i,t-1}$ depends on the size and sign of current and future shocks. In particular, our model enables specific shocks to erase the memory of past shocks. Furthermore, the interaction between the shock $\eta_{it}$ and the lagged persistent component $U_{i,t-1}$ is a central feature of our nonlinear approach. We then estimate the earnings model, and, given the estimated parameters, we simulate the model. \Cref{fig:simulatedY} shows that our nonlinear model reproduces the patterns of nonlinear persistence well.   

\begin{figure}[h!]
  \centering
 \includegraphics[width=.36\textwidth]{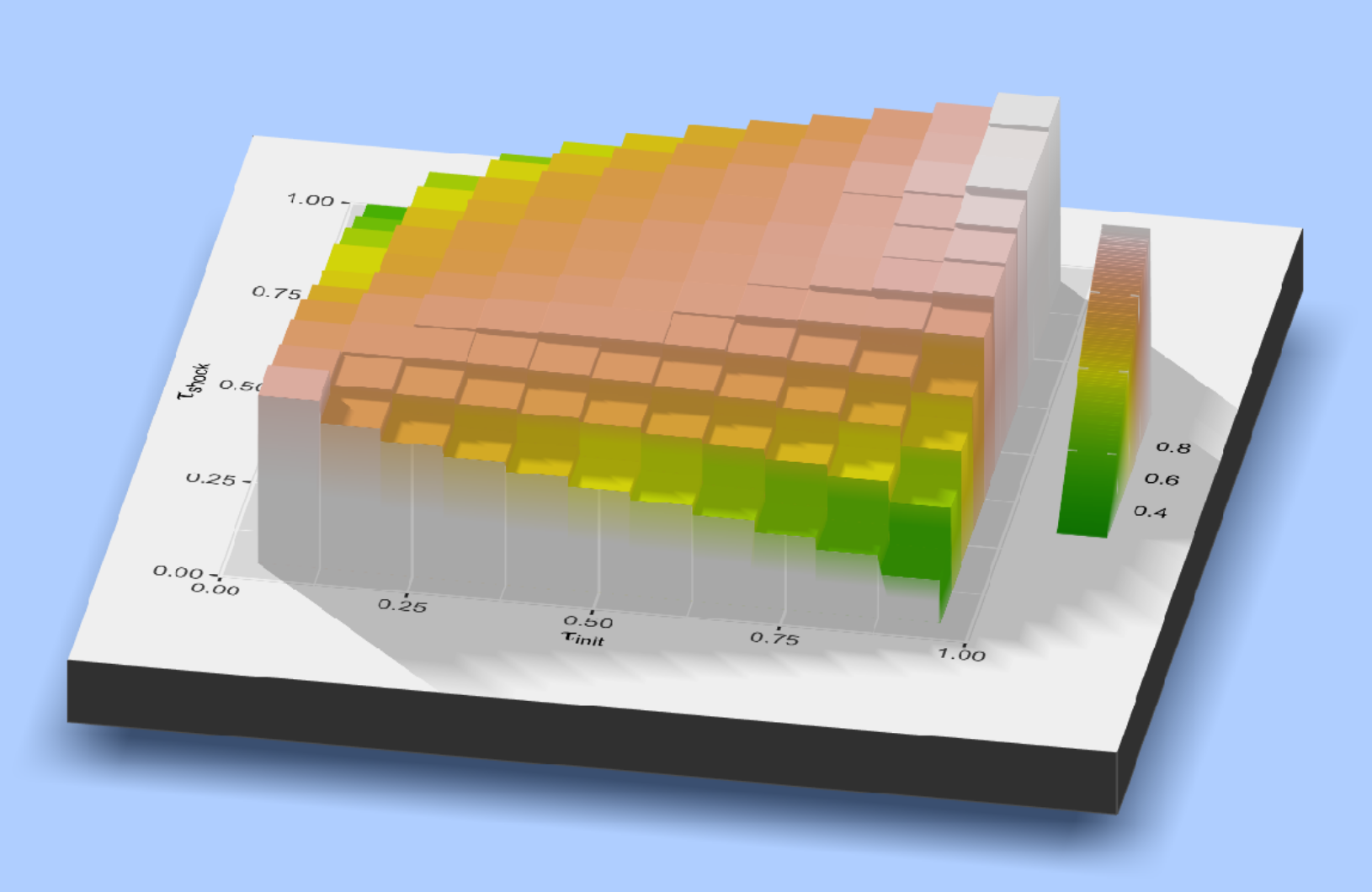}
 \caption{Estimates of the average derivative of the conditional quantile function of log-earnings residuals $y_{it}$ given $y_{i,t-1}$ with respect to $y_{i,t-1}$ in the PSID}
 \label{fig:realY}
\end{figure}

\begin{figure}[h!]
 \centering 
 \includegraphics[width=.36\textwidth]{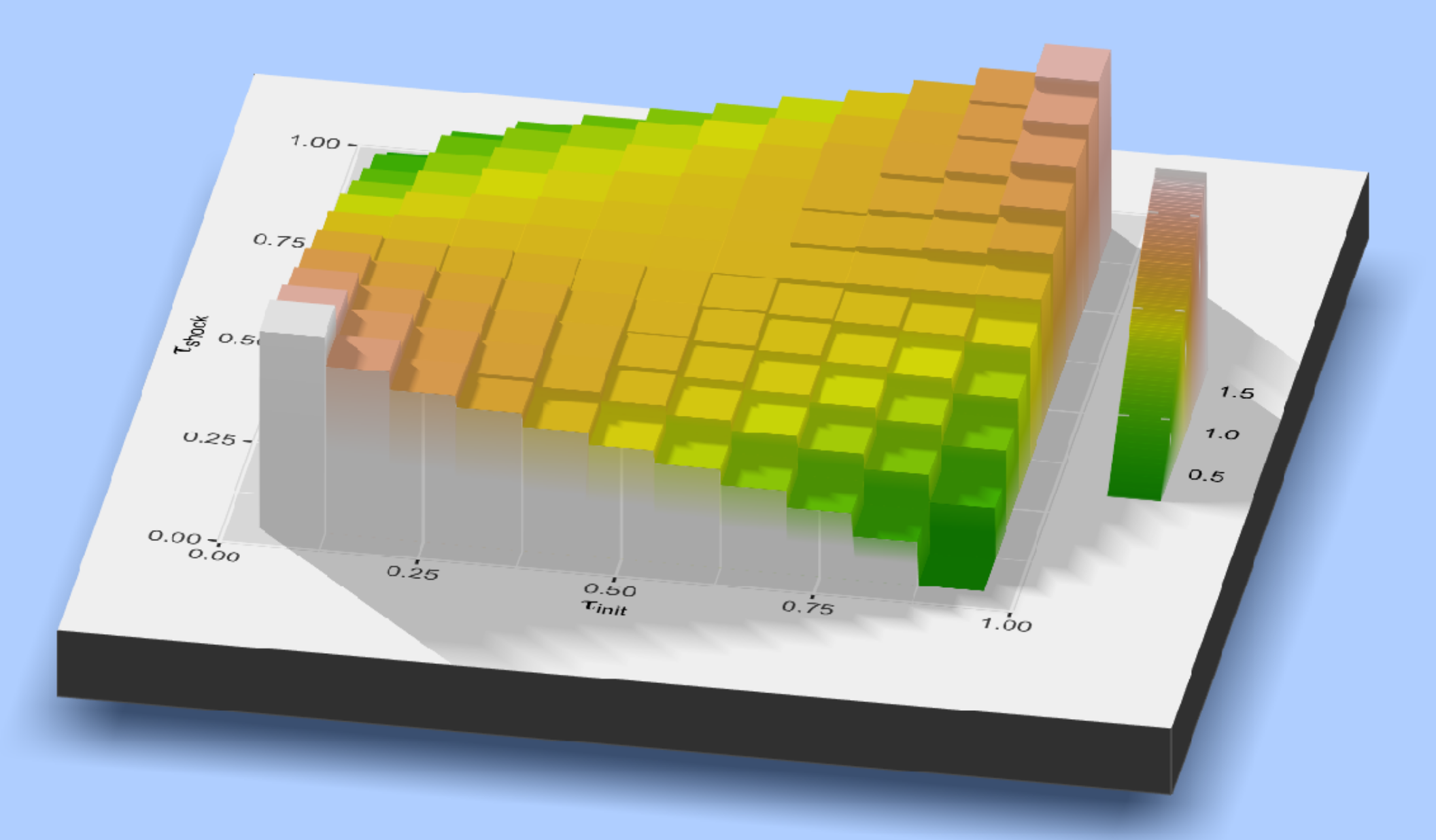}
 \caption{Estimates of the average derivative of the conditional quantile function of the persistent component $U$.}
 \label{fig:simulatedU}
\end{figure}

\Cref{fig:realY} indicates the presence of nonlinear persistence, which depends on both the percentile of past earnings $(\tau_{\textsf{init}})$ and the percentile of the quantile innovation $(\tau_{\textsf{shock}})$. \Cref{fig:simulatedU} then shows the estimated persistence of the earnings component $U_{it}$. Specifically, the graph shows the marginal effects, evaluated at percentiles $\tau_{\textsf{init}}$ and $\tau_{\textsf{shock}}$ and at the mean age in the sample. Persistence in $U$'s is higher than persistence in log-earnings residuals, consistently with the fact that \Cref{fig:simulatedU} is net of transitory shocks. One observation that sets this study apart from \cite{arellano2017} is that the persistence in \Cref{fig:simulatedU} is higher than 1. For high-earnings households hit by good shocks and low-earnings households hit by bad shocks, persistence is even above 1.5, with the persistence in the latter being higher than that in the former.

\begin{figure}[h!]
 \centering 
 \includegraphics[width=.36\textwidth]{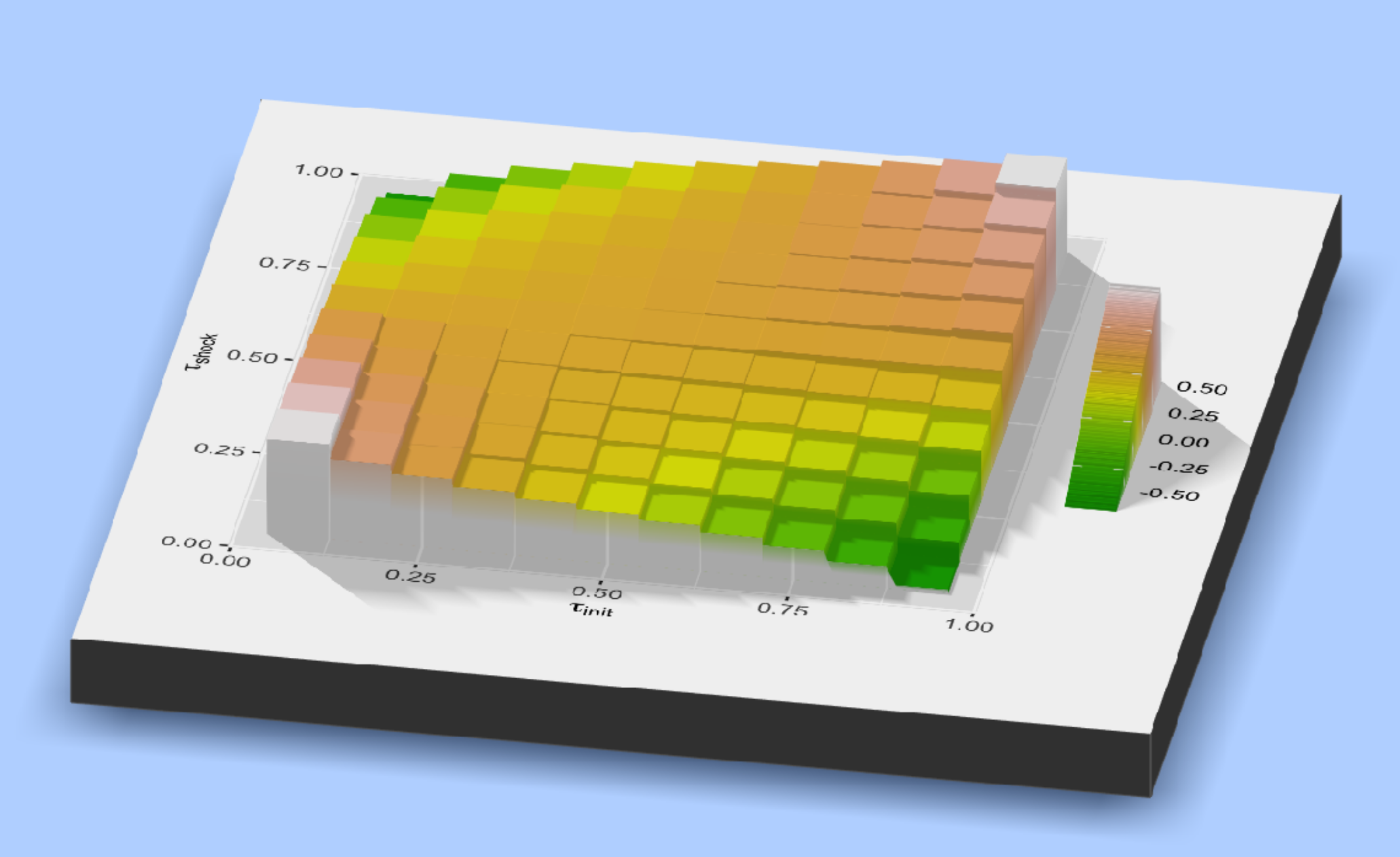}
 \caption{Estimates of the average derivative of the conditional quantile function of the transitory component $V$.}
 \label{fig:simulatedV}
\end{figure}

\begin{figure}[h!]
 \centering 
 \includegraphics[width=.36\textwidth]{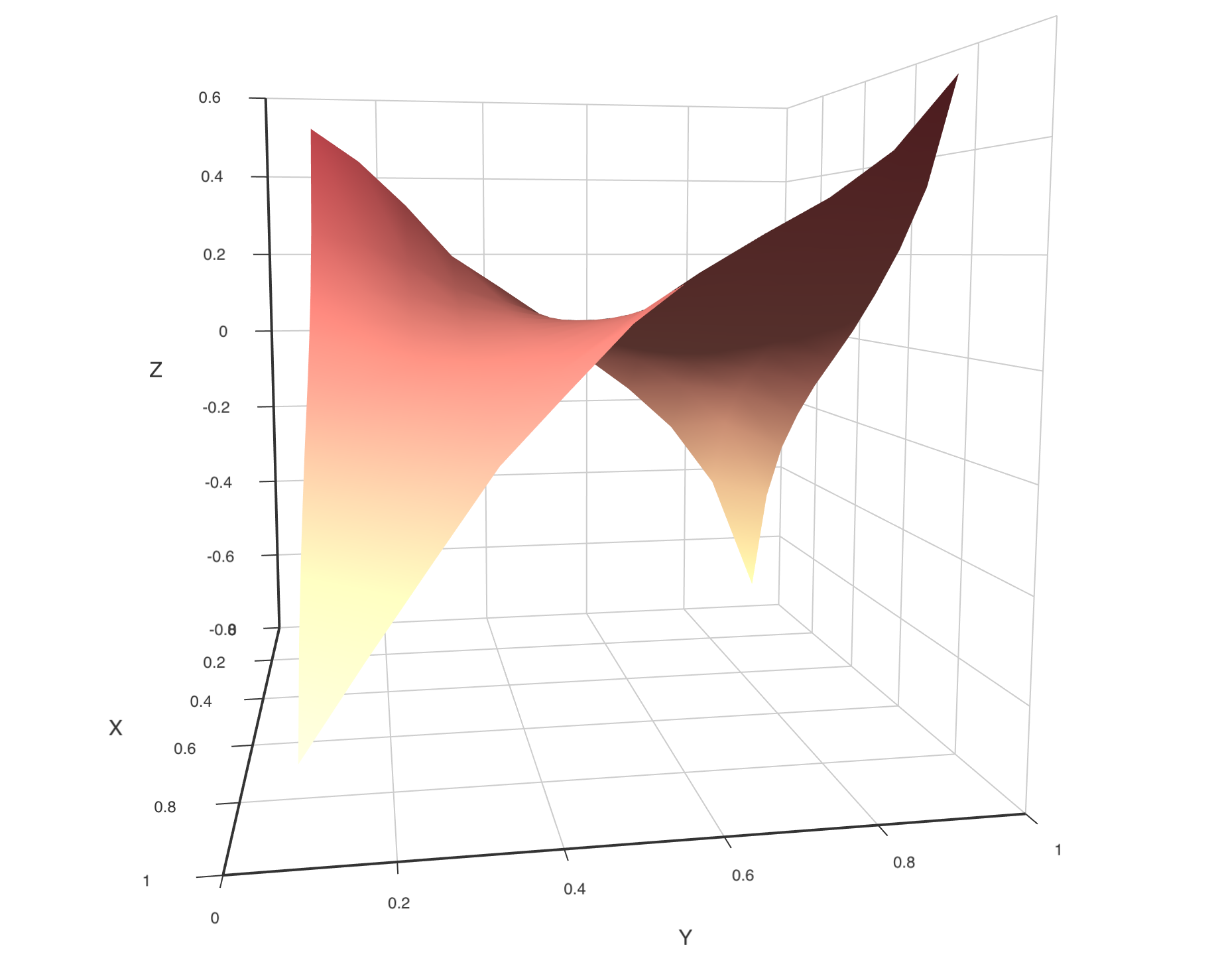}
 \caption{Estimates of the average derivative of the conditional quantile function of the persistent component $V$.}
 \label{fig:simulatedV_e4}
\end{figure}

\Cref{fig:simulatedV,fig:simulatedV_e4} demonstrate that the persistence in $V$'s is generally lower in  magnitude than that in $U$. One notable feature is that when high-earnings households are hit by bad shocks and low-earnings households are hit by good shock, the persistence can be negative. The high degree of nonlinearity displayed here strongly rejects that $V_t$ follows an independent process across $t$. If it did, its nonlinear persistence measures would remain constant for any $t$.  

\subsubsection{ARCH effects}
\begin{figure}[h]
  \centering
	\subfigure{       
      \begin{minipage}{2.5cm}                                                          \includegraphics[scale=0.14]{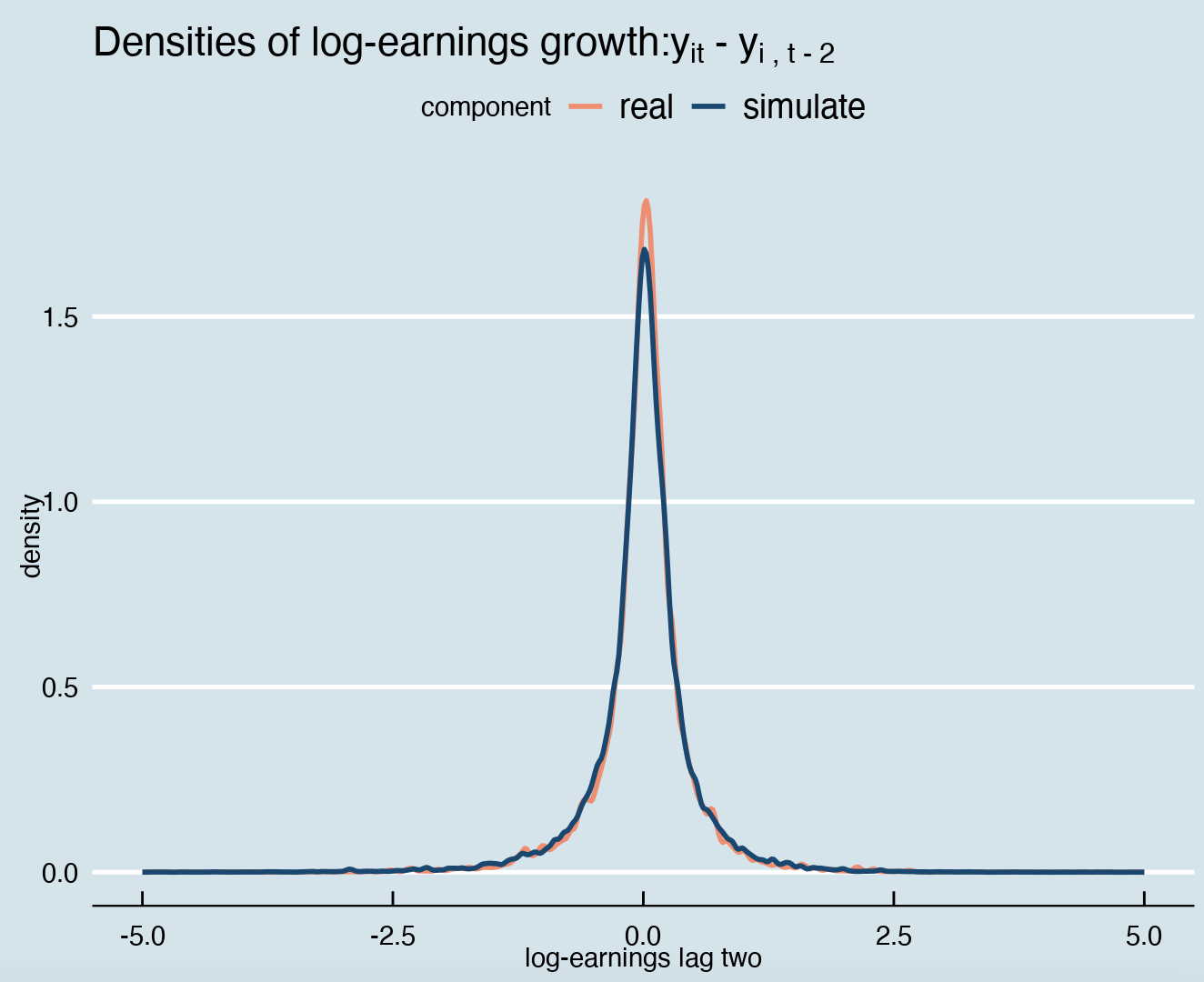} 
	\end{minipage}} 
    \hspace{1cm}
	\subfigure{  
      \begin{minipage}{2.5cm}                                                          \includegraphics[scale=0.14]{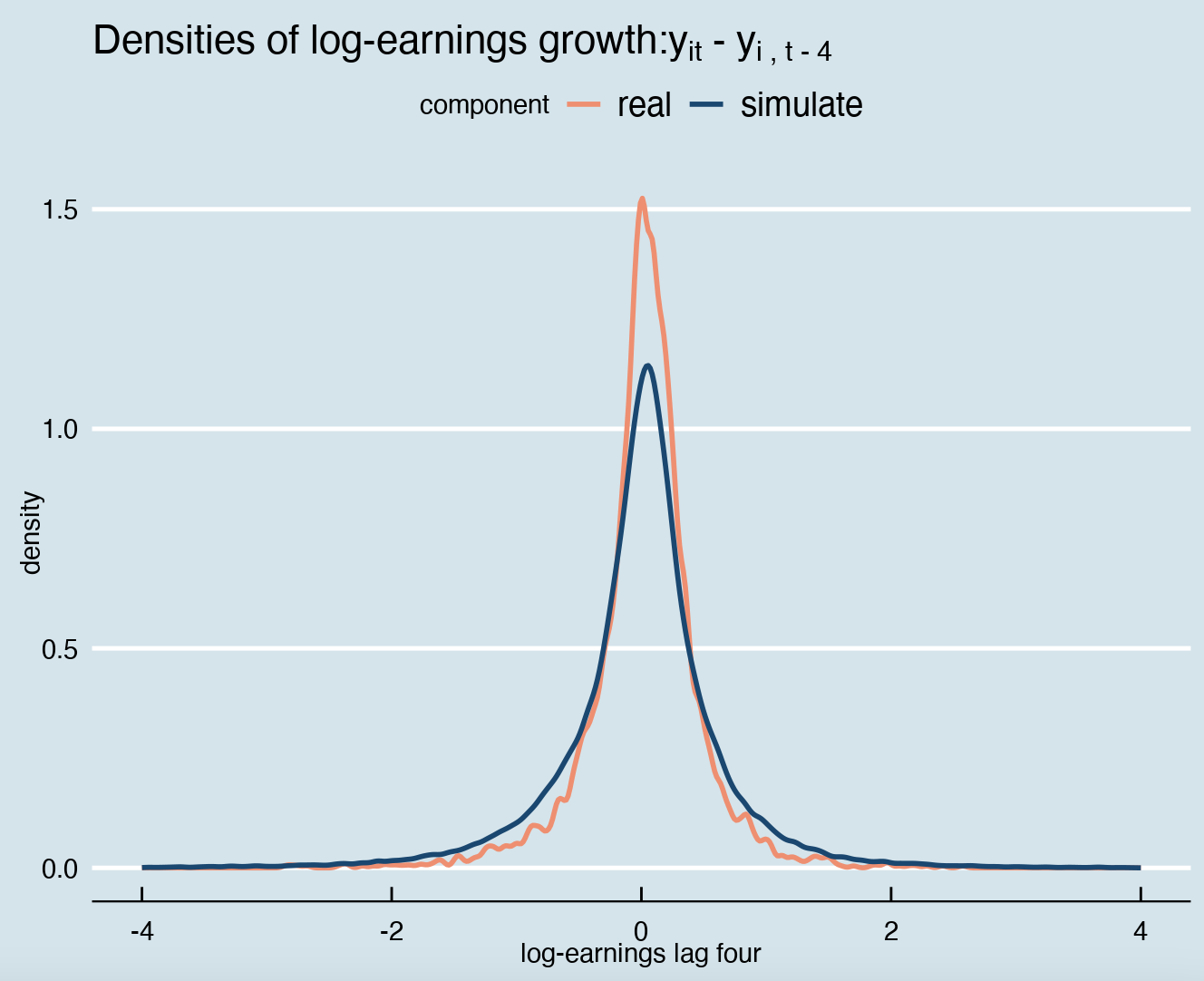}  
	\end{minipage}}  \\
	\subfigure{  
      \begin{minipage}{2.5cm}                                                          \includegraphics[scale=0.14]{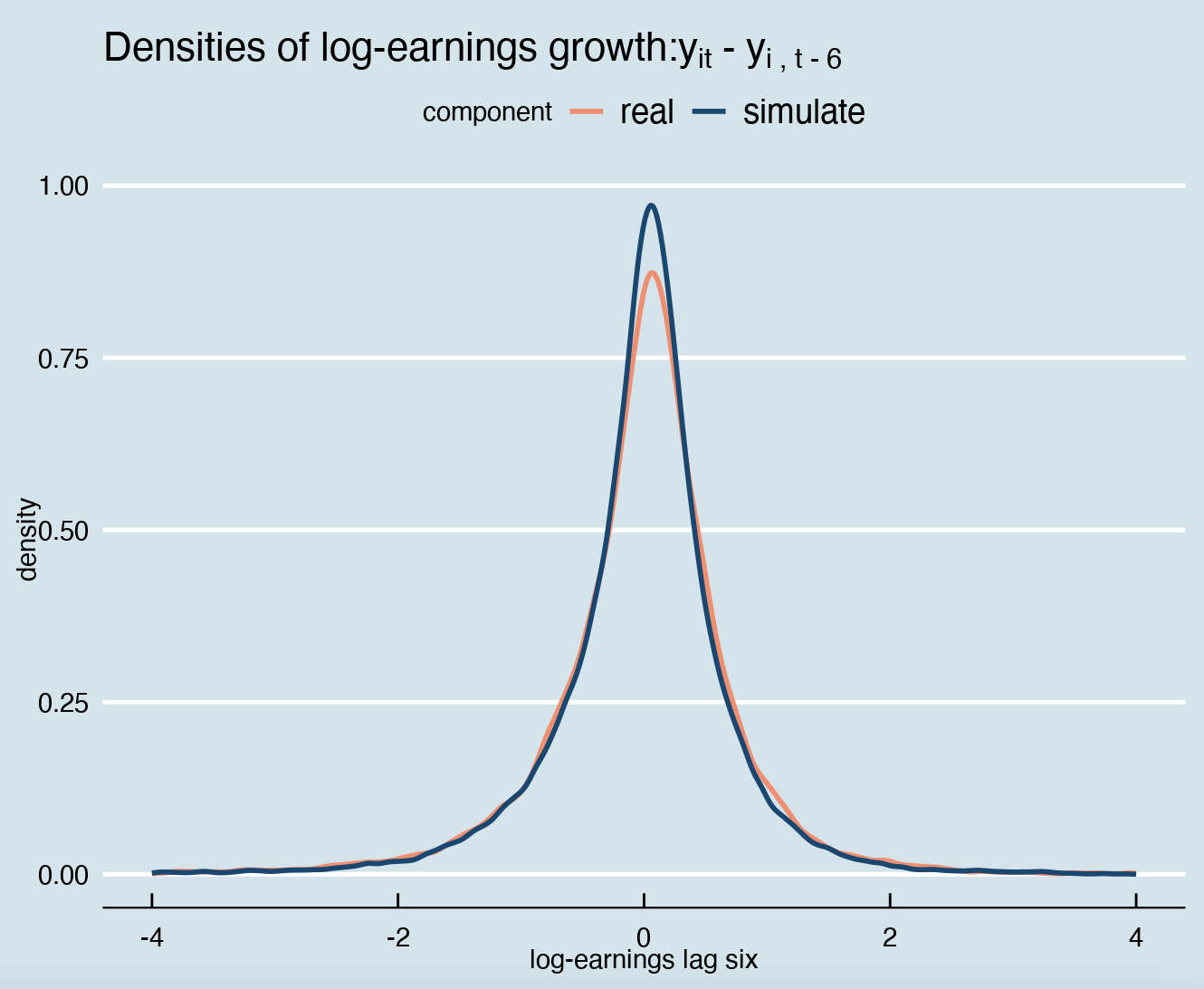}  
	\end{minipage}}  
    \hspace{1cm}
	\subfigure{  
      \begin{minipage}{2.5cm}                                                          \includegraphics[scale=0.14]{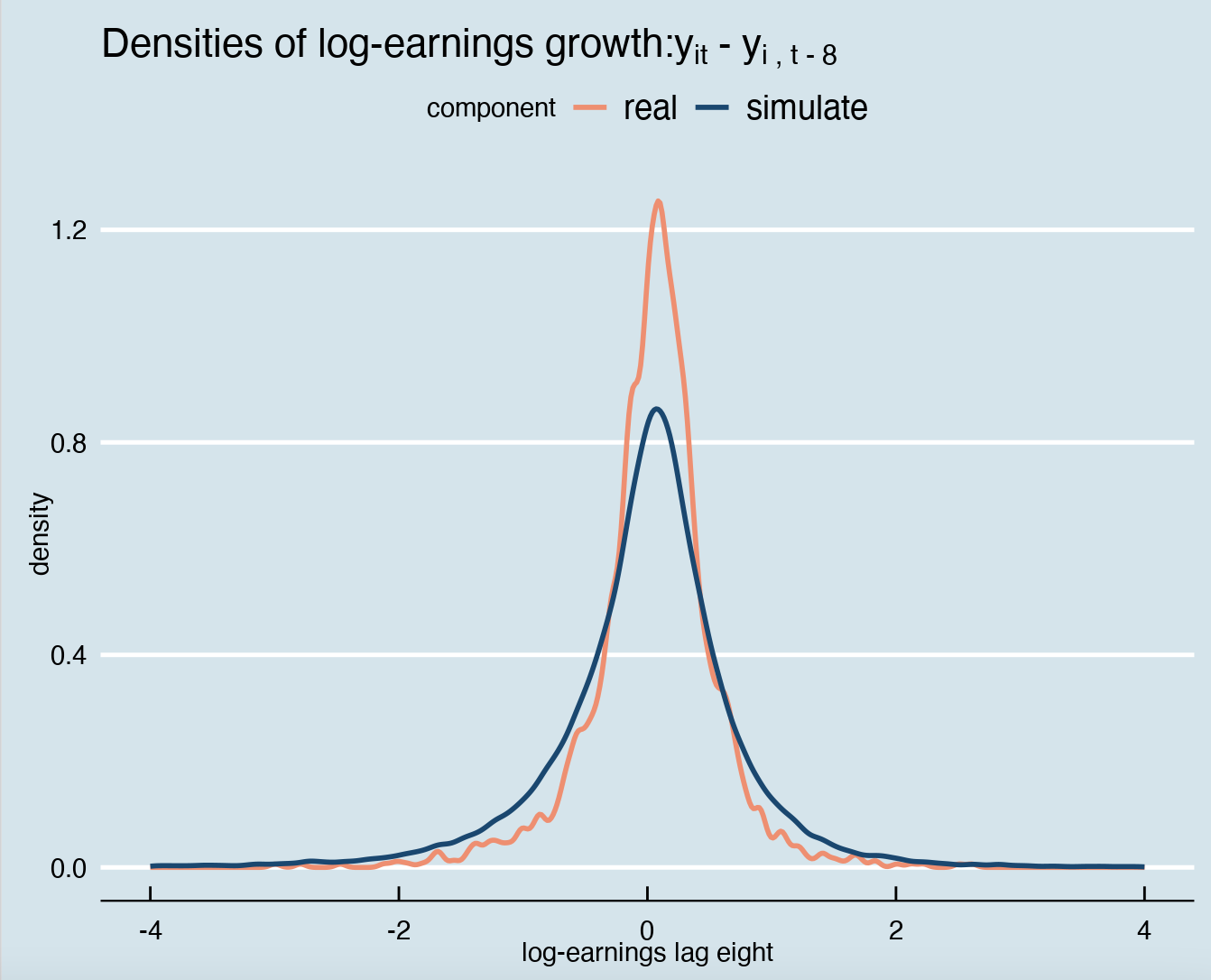}  
	\end{minipage}} 
	 \caption{Densities of log-earnings growth at various horizons.} 
	\label{fig:arch}  

\end{figure}

\Cref{fig:arch} presents estimates of log-earnings residuals growth at various horizons, from 2 to 8 years. All of them suggest the presence of ARCH effects, which is consistent with findings in the existing literature, such as \cite{Meghir2004}. The data also reveal that log-earnings growth is non-Gaussian and displays negative skewness and high kurtosis. \cite{Guvenen2015} finds similar features in U.S. administrative data. \cite{arellano2017} further highlights the sknewness and excess kurtosis of log-earnings growth at long horizons are primarily due to the non-Gaussianity of the transitory component. 

%

%


\section{Conclusion}

We develop a nonparametric identification strategy for modeling earnings dynamics, differentiating the two unobserved components  based on their distinct impact on household consumption. We also propose an modified stochastic EM algorithm for estimating this model. The identification tool relies on the assumptions that several linear operators are one-to-one.

In analyzing PSID, the empirical results reveal notable nonlinearities in both persistence component and transitory component. Specifically, substantial nonlinear persistence and conditional skewness are observed in both components. These findings suggest that the earnings shocks to a household depend on both the history of past shocks and the household's past relative wealth. In particular, persistence is higher for high-earnings households hit by good shocks and low-earnings households hit by bad shocks, while it is lower for high-earnings household hit by bad shocks and low-earnings households hit by good shocks.  These features align with similar observations in the PSID that previous earnings dynamic models cannot capture. We also find some other features such as ARCH effects that have been documented in other literature.


\section*{Acknowledgment}

The authors would like to thank the University of Michigan for providing the Panel Study of Income Dynamics (PSID) data, which was essential to the success of our research. We are also deeply grateful to Professor Roger Koenker for his valuable suggestions and clarifications on issues about quantile regression models.


%

\end{document}